\documentclass{cernrep}
\usepackage{graphicx,here}
 
\begin{document}

 \title{UPPER LIMITS IN THE CASE THAT ZERO EVENTS ARE OBSERVED:
AN INTUITIVE SOLUTION TO THE BACKGROUND DEPENDENCE PUZZLE}
\author{P. Astone, G. Pizzella}
 
\institute{Sezione INFN Roma 1 (``La Sapienza''), 
Universit{\`a} ``Tor Vergata'' and
LNF Frascati. Italy}

%
%
%

\maketitle 
 
\begin{abstract}
We compare the ``unified approach''
for the estimation of upper
limits with an approach based on the Bayes theory, in the
special case that no events are observed. 
The ``unified approach'' predicts, in this case, 
an upper limit that decreases with the increase
in the expected level of background. This seems absurd.
On the other hand, the Bayesian approach leads to a result 
which is background 
independent.
An explanation of the Bayesian result is presented, together with 
suggested reasons
for the paradoxical result of the ``unified approach''. 
\end{abstract}
\section{INTRODUCTION}
The study of a new phenomenon in science often ends up in a null result.
However it might be of great importance to set upper limits, as
this will help our understanding by eliminating some of the theories
proposed.

The determination of upper limits is presently a hotly debated issue in
several fields of physics. Many papers have been devoted 
to this problem and different solutions have been proposed. 
In particular the problem has been discussed 
in paper \cite{cousins} (``unified approach'') 
and, more recently, in papers \cite{ci,inferring},
based on the Bayes' theory.
The use of the ``unified approach'' (FC) to set upper limits
or confidence intervals is recommended by the PDG \cite{PDG}.
The ``unified'' and the Bayesian approaches are very different, not 
only in the sense that they lead to
different numerical results but more radically in the meaning they attribute
to the quantities involved. These differences lead to intrinsic problems
in any comparison of their separate results.
The purpose of this letter is to try to throw some light on this
contentious and important issue. We shall show that the Bayesian approach 
is the correct one. 
If our argument is accepted by the scientific community,
many debates about upper limits will be clarified.

\section{THE BACKGROUND DEPENDENCE PUZZLE}
According to the (FC) ``unified approach'' 
the upper limit is calculated using 
a revised version of the classical Neyman
construction for confidence intervals. This approach is usually referred 
to as the ``unified approach to the classical statistical analysis'', and
it aims to
unify the treatment of upper limits and confidence intervals.
On the Bayes side,
according to \cite{ci,inferring}, the upper limit may be calculated  using
a function ${\cal R}$ that is proportional to the likelihood.
This function is called the ''relative belief updating ratio''
and has already been used to analyse data in papers \cite{zeus,higgs}. 
The procedure has been
extensively described by G. D' Agostini in \cite{cern}.

Comparison between the two approaches is difficult 
for the general case. But we have noticed
a special case which is easier to discuss. In this case the
greater efficacy of one approach compared to the other one seems clear.
This case is when the experiment gave no
events, even in the presence of a background greater than zero.

When there are zero counts,
the predictions obtained with the two methods 
are different and both are -intuitively- quite disturbing.
Our intuition would, in fact, be satisfied by an upper limit that
increases with the background level, and this is, in general, the case when
the observation gives a number of events of the order of the background.
However, when zero events are observed, the 
``unified approach'' upper limit decreases if
the background increases (a 
noisier experiment puts a better
upper limit than a less noisy one, which seems 
absurd) while the Bayesian approach
leads to the predictions that a constant upper limit will be found
(the upper limit does not depend on the noise of the experiment).
Various papers\cite{giunti,punzi,woodroofe}
 have been devoted to the problem of solving some
intrinsic difficulties with the ''unified'' approach:
in particular to solving the problem of ''enhancing the physical significance
of frequentist confidence intervals''\cite{giunti}, or 
to imposing ''stronger classical confidence limits''\cite{punzi}. 
In this latter article the 
proposed method 
''gives limits that do not depend on background in the case of no
observed events'' (that is the Bayesian result !).

In what follows we will give an explanation for the two results.

We remind the reader that
the physical quantity for which a limit must be found is the events
rate (i.e. a gravitational wave burst rate) $r$.
Here we will assume  stationary working conditions.
For a given hypothesis $r$,  
the number of events which can 
be observed in the observation time $T$ is described by a Poisson 
process which has an intensity equal to the sum
of that due to background and that due to signal.

In general, the main ingredients in our problem are that:
\begin{itemize}

\item
we are practically sure about the expected rate of background 
events $r_b=n_b/T$ but not about the number of events that will 
actually be observed (which will depend on the Poissonian statistics).
 $T$ is the observation time;

\item
we have observed a number $n_c$  of events but,
obviously, we do not know how many of these events have to be
attributed to background and how many (if any) to true signals.
\end{itemize}

Under the stated assumptions, the likelihood is \begin{equation}
f(n_c\,|\,r,r_b) =\frac{e^{-(r+r_b)\,T}((r+r_b)\,T)^{n_c}}{n_c!}\,,
\label{eq:likr}
\end{equation}

We will now concentrate on the solution given by the Bayesian approach.

The  ``relative belief updating ratio'' ${\cal R}$ is defined as:

\begin{equation}
{\cal R}(r;n_c,r_b,T) = \frac{f(n_c\,|\,r,r_b)}{f(n_c\,|\,r=0,r_b)}\,,
\label{eq:rbur_def}
\end{equation}


This function is proportional to the likelihood and it allows us to
infer the probability that  $rT$ signals will be observed for given priors
(using the Bayes's theorem).

Under the hypothesis $r_b>0$ if  $n_c >0$,  ${\cal R}$ becomes 

\begin{equation}
{\cal R}(r;n_c,r_b,T) = e^{-r\,T}\left(1+\frac{r}{r_b}\right)^{n_c}\,.
\label{eq:erre}
\end{equation}

The upper limit, or -more properly- ''standard sensitivity bound" \cite{cern},
can then be calculated using the
 ${\cal R}$ function: it
is the value $r_{ssb}$ obtained when
 
\begin{equation}
{\cal R}(r_{ssb};n_c;r_b;T)=0.05
\label{cinque}
\end{equation}

We remark that 5\% does not represent a probability, but is a useful way
to put a limit independently of the priors.

Eq. \ref{eq:erre}
when no events are observed, that is, when $n_c$=0, becomes:
\begin{equation}
{\cal R}(r) = e^{-rT}\label{noevents}
\end{equation}

Thus putting $n_c=0$ in Eq. \ref{cinque} we
find $r_{ssb}=2.99$, independently of the value of the background
$n_b$.

We will not describe the well known  (FC) procedure here, but we
would just observe that, according to this procedure,
for $n_c=0$ and $n_b=0$, the upper limit is 3.09 (numerically almost identical
to the Bayes' one) $but$ it decreases as $n_b$ increases
(e.g. for $n_c=0$ and $n_b=15$ the upper (FC) limit at 95\% CL is 1.47).

In an attempt to
understand such different behaviour we will now discuss some
particular cases.
Suppose we have $n_c=0$ and $n_b\not=0$. This certainly means that
the number of accidentals, whose average value can be determined
with any desired accuracy, has undergone a fluctuation.
The larger the $n_b$ values, the smaller is the $a~priori$ probability
that such fluctuations will occur.
 Thus one could reason that it is less likely that a
number $n_{gw}$ of real signals could have been 
associated with a large value of  $n_b$, since the
observation gave $n_c=0$. 

According to the Bayesian approach, instead, one cannot
ignore the fact that the observation $n_c=0$ has already being made at the time
the estimation of the upper limit comes to be calculated. 
The Bayesian  approach requires
that, given $n_c=0$ and $n_b\not=0$, one 
evaluates the $chance$ that  a number
$n_{gw}$ of signals exists. This $chance$ of a possible signal
is applied to  the observation that has already been
made.

Suppose that we have estimated the average background
with a high degree of accuracy, for example $n_b$=10.
In the absence of signals,
the a priori probability of observing zero events, due just
 to a background fluctuation, is given by 
\begin{equation}
f_{n}=f(n_c=0 | n_b=10)=e^{-n_b}=4.5 \cdot 10^{-5}
\label{solonoise}
\end{equation}
 
Now, suppose that we have measured zero events, that is $n_c$=0. 
In general $n_c=(n_b+n_{gw})$.
It is now nonsense to ask what the probability that $n_c$=0 is, 
since the experiment has already been made and the probability is 1.

We may ask how the a priori probability 
would be changed if $n_{gw}$ signals were added to
the background. We get
\begin{equation}
 f_{sn}=f(n_c=0 |n_b=10,n_{gw})=e^{-(n_b+n_{gw})}
\label{signalnoise}
\end{equation}

It is obvious that $f_{sn}$ can only decrease relative to $f_{n}$, since we are
considering models in which signal events can only
add to noise events\footnote{
In a gravitational wave experiment signals may add up to the noise
with the same phase, thus increasing the energy of the combined effect, 
or with a phase opposite to that of the noise, thus reducing the
energy.
They can in particular add up also to noise events, even if we expect this
to happen with a very low probability, as we know that the events due to
the signal are very ``rare'' compared to the events due to the noise.

Anyway, in principle, the presence of this fact will lead to the prediction of
a signal rate that increases with the background: in fact  
the probability that one background event be cancelled by a 
signal event increases, as $n_b$ 
increases. Thus, if we, at least in part, attribute the observation
of $n_c$=0 to a cancellation of background events due to
the signal the final limit on $r$ should increase.

In the modelling we usually, as reasonable, consider this effect be
negligible. If this is not the case then it must be properly modelled
in the likelihood.}.

The right answer is guaranteed if the question is well posed.
Given all the previous comments, the most obvious question at this point is:
{\bf what is that signal $n_{gw}$
which would have reduced the probability $f_n$
by a constant factor, for example 0.05 ? }

\begin{equation}
f_{sn}=f_{n} \cdot 0.05=e^{-n_b}\cdot e^{-n_{gw}}
\label{finale}
\end{equation}

Using Eqs. \ref{solonoise}, \ref{signalnoise} and
\ref{finale} the solution is:
\begin{equation}
e^{-n_{gw}}=0.05  
\end{equation}
that is:
\begin{equation}
n_{gw}=2.99
\end{equation}

Now suppose another situation, $n_b$=20, thus 
$f_n=2.1 \cdot 10^{-9}$.
Repeating the previous reasoning we still get the limit 2.99.

The meaning of the Bayesian result is now clear:
we do not care about the absolute value
of the a priori probability of getting $n_c=0$ in the presence of noise alone.
The observation of $n_c=0$  means that the background gave
zero counts by chance. 
Even if the a priori probability 
is very small,
its value has no meaning once it has happened.
The fact that the single background  measurement 
turned out to be zero, either due to a
zero average background or due to the 
observation of a low (a priori) probability event,
must not change our prediction  concerning possible signals.

For $n_c=0$ we are certain that the number of events due to
the background is zero. Clearly this particular situation gives more
information about the possible signals.
In the case $n_c \not=0$, instead, it is not possible to distinguish 
between background and signal.
The mathematical aspect of this is that the Poisson 
formula when $n_c=0$ reduces
to the exponential term only, and thus it is possible to separate the two
contributions, of the signal (unknown) and of the noise (known).

We note that   
the different behaviour of the limit in the unified approach
is due to the non-Bayesian character of  the reasoning.
In such an approach an event that has already occurred is
considered ``improbable'': given the
observation of $n_c=0$  
they still consider that the probability
\begin{equation}
f_{sn}=f(n_c=0 |n_b,n_{gw})=e^{-(n_b+n_{gw})}\label{eq:fsn}
\end{equation}
decreases as $n_b$ increases. 
As a consequence they deduce that to a larger $n_b$ corresponds a smaller
upper limit $n_{gw}$.

Given the previous considerations, we must now 
admit that our intuition to expect an upper limit that increases with 
increasing background, even when $n_c=0$, was wrong. We should have 
expected to predict a constant signal rate, as a consequence of the
observation of zero events, independently of the background level.

\section{CONCLUSION}
We have compared the upper limits obtained with the (FC) ``unified'' and
with the Bayesian procedures,
in the case of zero observed events. 

We believe that the greater efficacy of the Bayesian approach 
compared to the (FC)
method, demonstrated for the case $n_c=0$, 
is a strong indication that the Bayesian method -natural, simple and intuitive-
is the correct one. Thus
we agree with the proposal in\cite{cern} that this method should  
be adopted by the scientific community
for upper limit calculations (see, for example,
\cite{nostroupper} on upper
limits in gravitational wave experiments).

\end{document}